\documentclass[floats,prb]{revtex4}
\usepackage{amsfonts}
\usepackage{bm}
\usepackage[latin9]{inputenc}
\usepackage{color}
\usepackage{verbatim}
\usepackage{amssymb}
\usepackage{graphicx}
\usepackage{esint}
\usepackage{times}

\begin{document}

\title{Weyl semimetals in optical lattices: moving and merging of Weyl points, and  hidden symmetry at  Weyl points}
\author{Jing-Min Hou$^1$  and Wei Chen$^2$}
\affiliation{$^1$Department of Physics,
Southeast University, Nanjing  211189, China\\
$^2$College of Science, Nanjing University of Aeronautics and
Astronautics, Nanjing 210016, China\\
\\
 Correspondence and requests for materials should be addressed to
J.-M.H. (e-mail: jmhou@seu.edu.cn)  }

\begin{abstract}
 We propose to realize Weyl semimetals in a cubic optical lattice. We find that there exist three distinct Weyl semimetal phases in the cubic optical lattice for different parameter ranges. One of them has two pairs of Weyl points and the other two have one pair of Weyl points in the Brillouin zone.  For a slab geometry with (010) surfaces, the Fermi arcs connecting the projections of Weyl points with opposite topological charges on the surface Brillouin zone is presented.  By adjusting the parameters,  the Weyl points can move in the Brillouin zone. Interestingly, for two pairs of Weyl points, as one pair of them meet and annihilate, the originial two Fermi arcs coneect into one. As the remaining Weyl points annihilate further, the Fermi arc vanishes and a gap is opened. Furthermore, we find that there always exists a hidden symmetry at Weyl points, regardless of  anywhere they located in the Brillouin zone.
 The hidden symmetry has an antiunitary operator with its square being $-1$.
\end{abstract}

\maketitle

 In last decade, topological matters  have   become an
 important branch of condensed
matter physics\cite{Hasan10,Qi11}.  Previously, the studies mainly concentrate on gapped systems, such as integer quantum Hall insualtors\cite{Thouless} and quantum anomalous Hall insulator\cite{Haldane}, topological insulators\cite{Kane}, chiral
topological superfluids\cite{Read},   helical topological
superfluids or superconductors\cite{Qi2}, and so on. Recently, physicists pay much attention on the topological characters of gapless systems, which were dubbed as topological semimetals.
Generally, topological semimetals include Weyl semimetals\cite{Wan,Xu,Burkov1,Burkov2,Zyuzin}, Dirac semimetals\cite{Yang14,Xu1}, topological nodal-line semimetals\cite{Burkov1}.
For Weyl  semimetals, the materials have
band structures with band-touching nodal points  in momentum space, where the isolated band degeneracy occurs.  Near these touching points, the dispersion relation
is linear and can be described by a massless two-component Weyl
Hamiltonian. At the nodal points, there exist singularities of a Berry field. Integrating the Berry field on the surface enclosing
the singular point in momentum space, one obtain a Chern number, i.e., a topological charge. Thus, the band-touching nodal points can be considered as   monopoles in momentum space. Due to the Nielsen-Ninomya theorem, the nodal points with opposite topological charges appear in pairs. The meeting of opposite charges in momentum space can lead to annihilation of nodal point pairs. The opposite topological charges  can be  separated from each other in
momentum space so that they  cannot be destroyed   by the mutual annihilation if the time-reversal symmetry or inversion symmetry is broken.    Time reversal symmetry breaking Weyl semimetals were firstly predicted in pyrochlore iridates\cite{Wan} and HgCr$_2$Se$_4$\cite{Xu}. Recently, inversion symmetry breaking Weyl semimetals were discovered in TaAs family\cite{Weng,Huang,Xu2,Lv,Xu3}. In the theoretical aspect, recently, Ganeshan and Das Sarma presented  a method to construct a Weyl semimetal by stacking one-dimensional Aubry-Andre-Harper lattice with tight-binding models with nontrivial topology\cite{Ganeshan15}, which provides a theoretical connection between the commensurate Aubry-Andre-Harper model in one dimension and Weyl semimetals in three dimensions.

It is a difficult task to investigate moving and merging of Weyl points and topological phase transitions in real solid materials, the parameters of which can not be tuned in a wide ranges. Fortunately, the
high controllability and tunability,  and large number of mature detection techniques
of cold atoms in optical lattices make them a
platform to stimulate
many interesting physics in condensed matters. Therefore, it is intriguing to study moving and merging of Weyl points, and topological phase transitions in optical lattices.
 In recent years, many schemes have
been proposed to realize various topological semimetals with neutral
atoms in optical lattices.
In two dimensions, gapless topological phases were proposed  in honeycomb optical lattices\cite{Zhu,Wunsch} and square optical lattices\cite{Hou09,Hou1,Hou2}. The important progress is the realization of topological semimetals in honeycomb optical lattices\cite{Tarruell}.
In three dimensions, Weyl semimetal were proposed to realized in optical lattices\cite{Delplace,Jiang,Dubcek,XuY}. In order to engineer the topological phases in optical lattices, sometimes, the hopping-accompanying phase, i.e., the Peirls phase, is required.  In experiments, the hopping-accompanying phase has been realized with periodic lattice shaking\cite{Struck12,Struck13} and laser-assisted tunneling techniques\cite{Aidelsburger11,Miyake13,Aidelsburger13,Jimenez-Garcia12}. Another important progress in experiments is the measurement of Zak phase of topological Bloch bands in optical lattices\cite{Atala13}, which provides a path to detect topological characters in optical lattices.

In this paper, we design a cubic optical lattice trapping cold fermionic atoms, which can be realized based the laser-tunnelling technique\cite{Aidelsburger11,Miyake13,Aidelsburger13,Jimenez-Garcia12}. In   different parameter ranges,
  the system supports three classes of Weyl semimetals, one of which has two pairs of  Weyl points in the Brillouin zone, the other two have  one pair of Weyl points in the Brillouin zone. By adjusting the parameters, we can study the moving and merging of Weyl points. When Weyl  points with opposite topological charges meet together, they annihilate and a topological phase transition happens. We also investigate the Fermi arc of surface states of a (010) slab. Fermi arcs connect the projections of Weyl poionts on the surface Brillouin zone and evolve with the moving of Weyl points. For the Weyl semimetal phase with two pairs of Weyl points, there are two Fermi arcs connect projections of Weyl points with opposite charges on the surface Brillouin zone. When a pair of Weyl points annihilate, the two Fermi arcs link into one single Fermi arc connecting the projections of the remaining Weyl points.  We find that the band degeneracy at Weyl points implies a hidden symmetry that has an antiunitary operator with its square being $-1$.
  Based on a mapping method, we discover the hidden symmetry at each Weyl point in the Brillouin zone and discuss its relation with topological phase  transitions.

\vspace{0.5cm} \noindent \textbf{Results} \\

\vspace{0.5cm} \noindent \textbf{Weyl semimetals in optical lattices}.

\begin{figure}[ht]
\begin{center}
\includegraphics[width=0.9\columnwidth]{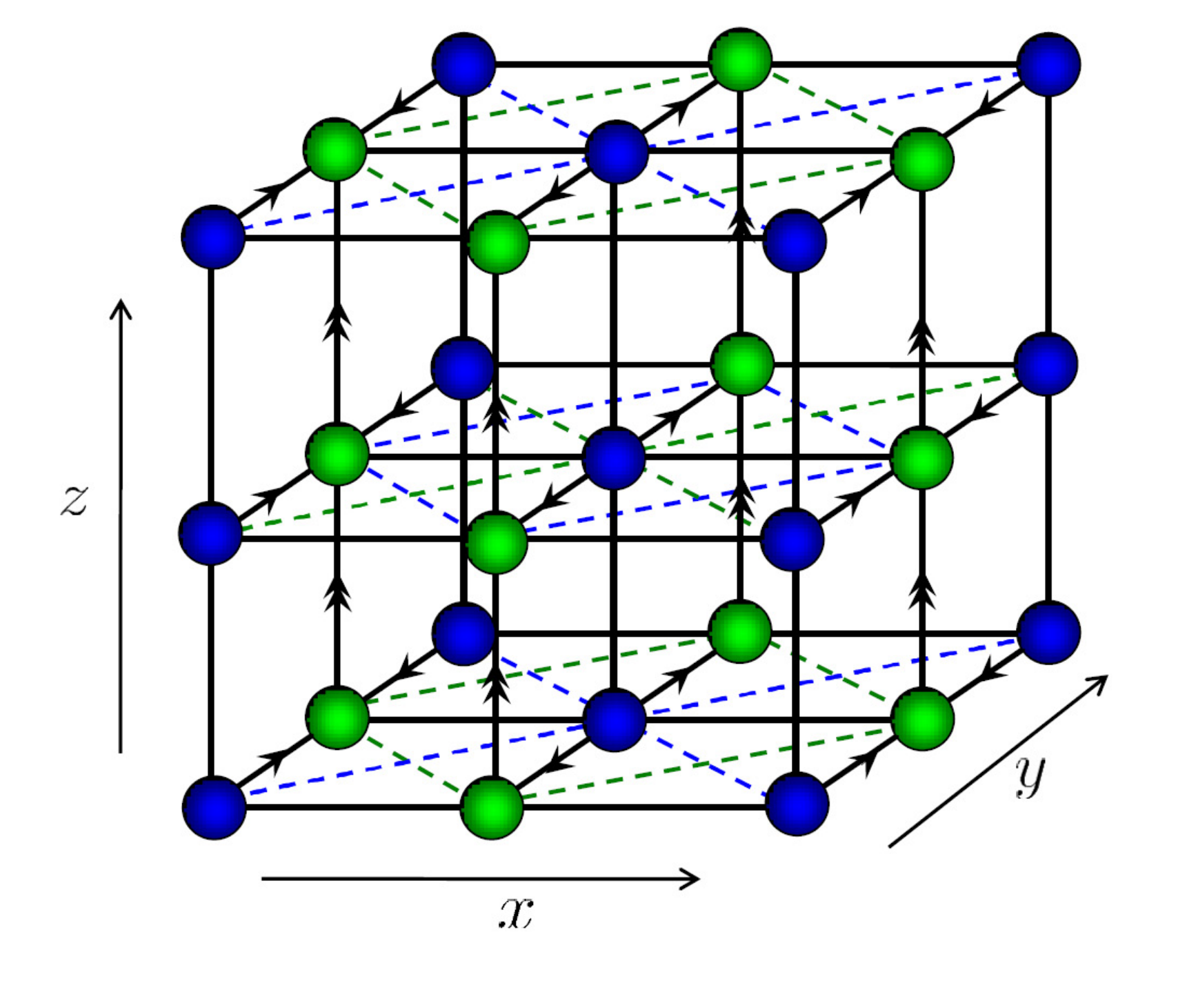}
\end{center}
\caption{\textbf{Schematic of the cubic optical lattice.} Here,
the blue and green balls represent sublattices A and B,
respectively; the single arrows and double arrows denote $\pi/2$ and
$\pi$ phases along with the hopping, respectively. }\label{fig1}
\end{figure}

 Here, we
consider a cubic optical lattice as shown in Fig.\ref{fig1}, where the
arrows represent the hopping-accompanying phase. The
hopping-accompanying phase is $\pi/2$ for the hopping along the $y$
axis and $\pi$ for the $z$ axis. Due to the appearing of the
hopping-accompanying phases, the translation symmetry is broken.
Thus the lattice is divided into two sublattices, i.e. sublattices
$A$ and $B$. Assuming the distance between the nearest lattice sites
being $1$, we define the primitive lattice vectors as
$\mathbf{a}_1=(1,-1,0)$, $\mathbf{a}_2=(1,1,0)$,
 and $\mathbf{a}_3=(0,0,1)$. The primitive reciprocal lattice
 vectors are
$\mathbf{b}_1=(\pi,-\pi,0)$, $\mathbf{b}_2=(\pi,\pi, 0)$, and
$\mathbf{b}_3=(0,0,2\pi)$.
Besides the hopping between nearest lattice sites,
  we also consider   the diagonal
hopping in the $x-y$ plane and a staggered potential. The corresponding Hamiltonian is $H=H_0+H_d+H_s$ with
\begin{eqnarray}
  H_0&=&-\sum_{i\in A} [t_x\hat a^\dag_{i}\hat
b_{i+\hat{x}}+t_x\hat a^\dag_{i}\hat
b_{i-\hat{x}}+t_ye^{-i\pi/2} \hat a^\dag_{i}\hat b_{i+\hat{y}}+t_ye^{-i\pi/2} \hat a^\dag_{i}\hat b_{i-\hat{y}}\nonumber\\
&&+t_z\hat a^\dag_{i}\hat a_{i+\hat{z}}+t_ze^{-i\pi}\hat
b^\dag_{i+\hat{x}}\hat b_{i+\hat{x}+\hat{z}}] +H.c., \label{tbh}
\end{eqnarray}
and
\begin{eqnarray}
H_d&=& -t_{xy}\sum_{i\in A}(a^\dag_i a_{i+\hat{x}+\hat{y}}-a^\dag_i
a_{i+\hat{x}-\hat{y}})\nonumber\\
&&+t_{xy}\sum_{j\in B}(b^\dag_j b_{j+\hat{x}+\hat{y}}-b^\dag_j
b_{j+\hat{x}-\hat{y}})+H.c.,\label{Hd}
\end{eqnarray}
and
\begin{eqnarray}
H_s=v\sum_{i\in A} a_i^\dag a_i-v\sum_{j\in B} b_j^\dag b_j,\label{Hs}
\end{eqnarray}
where $a_i$ and $b_i$ are the annihilation operators destructing a
particle at a lattice site of sublattice $A$ and $B$, respectively;
$t_x$ and $t_y$ represent the amplitudes of hopping along the $x$
and $y$ directions, respectively; $t_{xy}$ denotes the amplitude of hopping along the diagonal
direction; $v$ represents the magnitude of the staggered on-site
potential. This optical lattice can be realized through the laser-assisted tunneling technique, which has been applied in several experiments\cite{Aidelsburger11,Miyake13,Aidelsburger13,Jimenez-Garcia12}.

\begin{figure}

\includegraphics[width=1\columnwidth]{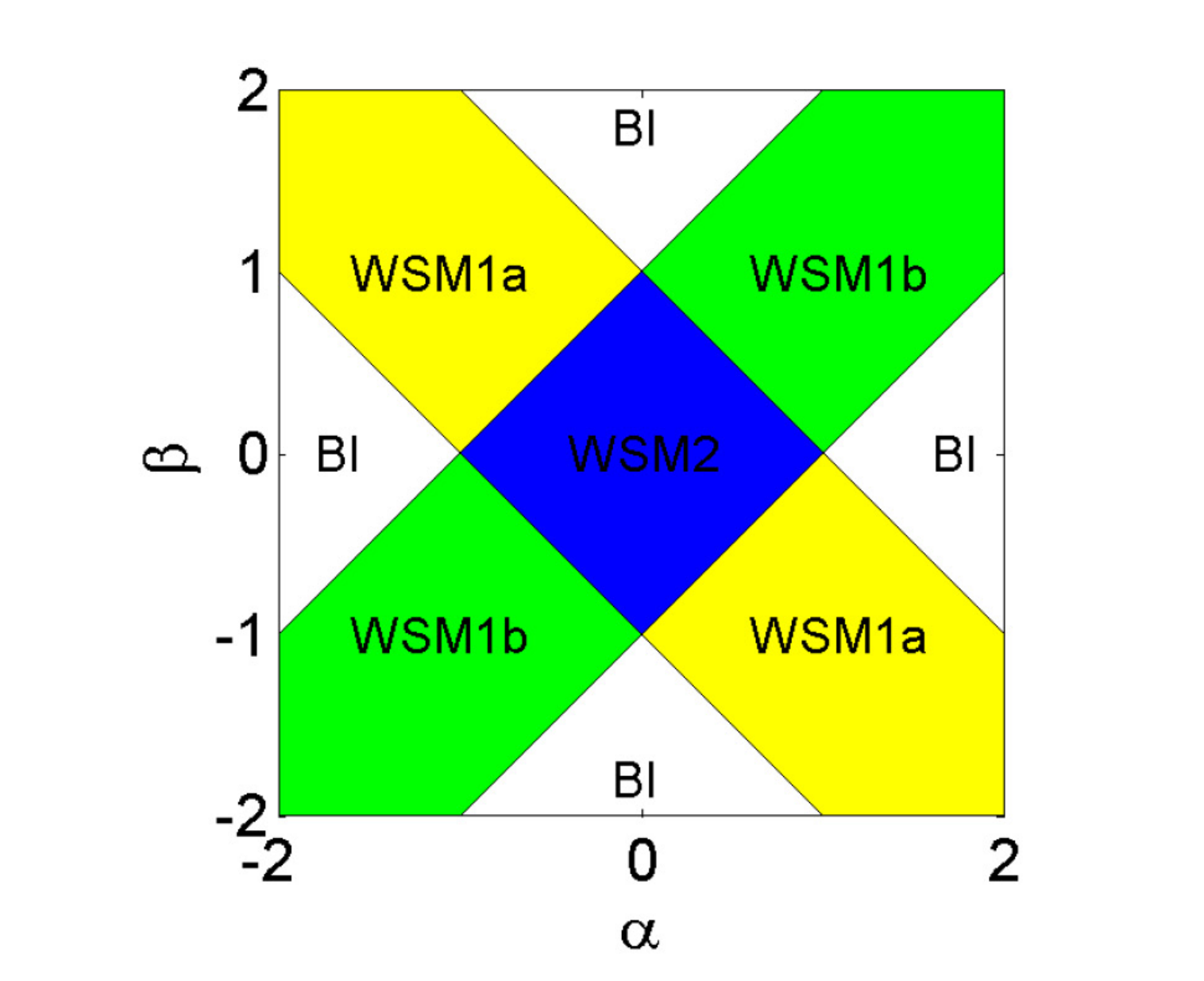}

\caption{\textbf{Schematic of the phase diagram.} Here, WSM2
(blue) denotes the Weyl semimetal phase with two pairs of Weyl
points in the Brillouin zone; WSW1a (yellow) and WSW1b (green)
denote the two semimetal phases with a pair of Weyl  points in the
Brillouin zone; BI denotes the band insulator phase.}\label{phase}
\end{figure}

Taking the Fourier's
transformation on equations (\ref{tbh}), (\ref{Hd}) and (\ref{Hs}), we rewritten the Hamiltonian
as $H=[a^\dag_\mathbf{k},
b^\dag_\mathbf{k}]^T\mathcal{H}(\mathbf{k})[a_\mathbf{k},b_\mathbf{k}]$, where $\mathcal{H}(\mathbf{k})$ is the corresponding Bloch Hamiltonian   as
\begin{eqnarray}
\mathcal{H}(\mathbf{k})&=&-2t_x\cos k_x \sigma_x-2t_y\cos k_y \sigma_y\nonumber\\
&&-2t_z(\cos k_z-\alpha-\beta\sin k_x \sin k_y)\sigma_z, \label{BH1}
\end{eqnarray}
 with $\alpha=v/2t_z$ and $\beta=2t_{xy}/t_z$ being the
dimensionless parameters and $\sigma_{x}, \sigma_y$ and $\sigma_z$ being the Pauli
matrices defined in the sublattice space. Diagonalizing equation
(\ref{BH1}), we obtain the corresponding dispersion relation as
\begin{eqnarray}
E=\pm \sqrt{4t_x^2\cos^2k_x+4t_y^2\cos^2k_y+(2t_z\cos k_z-m )^2},
\end{eqnarray}
with $m=2t_z(\alpha+\beta\sin k_x \sin k_y)$.
From this dispersion relation,  we can see that two band touch at some points $\mathbf{W}_i$ in the Brillouin zone in some parameter ranges.
Near the touching points, the dispersion relation has the linear
form as
\begin{eqnarray}
h(\mathbf{p})=v_x p_x\sigma_x\pm v_yp_y\sigma_y\pm v_zp_z\sigma_z,
\end{eqnarray}
with $\mathbf{p}=\mathbf{k}-\mathbf{W}_i$.  Around the the touching
points, the chirality can defined as
\begin{eqnarray}
C=\textrm{sgn}[\det(v_{ij})]=\pm 1. \label{W}
\end{eqnarray}
which is also the topological charge at Weyl points.
Thus, the touching points are Weyl points and,
correspondingly, the system is a Weyl semimetal phase.
According to the number of Weyl points in different parameter ranges, we can classify the system into four phases: (i) When  $\left|\alpha+\beta\right|<1$ and $\left|\alpha-\beta\right|<1$ are satisfied, there are
four distinct points $\mathbf{W}_{1,2}=(\pi/2,\pi/2, \pm \arccos(\alpha+\beta))$ and
$\mathbf{W}_{3,4}=(\pi/2, -\pi/2, \pm \arccos(\alpha-\beta))$ in the Brillouin zone.
Since there
are two pairs of Weyl points in the Brillouin zone, we term this
phase as   WSM2 phase.
(ii) For the case
$\left|\alpha+\beta\right|<1$ and $\left|\alpha-\beta\right|>1$,  only the pair
$\mathbf{W}_{1,2}$ exists.
  Thus, we term this   phase as WSM1a phase.
(iii) For the case
$\left|\alpha+\beta\right|>1$ and $\left|\alpha-\beta\right|<1$,
where the Weyl points $\mathbf{W}_{3,4}$ still remain. We term this
new phase as WSM1b phase, which is different from the WSM1a phase. (iv)
For $\left|\alpha+\beta\right|>1$ and $\left|\alpha-\beta\right|>1$,
no Weyl point exists and a gap opens, so the system
is a band insulator. The phase diagram  is shown as in
Fig.\ref{phase}.

\vspace{0.5cm} \noindent \textbf{Moving and merging of Weyl points, topological phase transition, and Fermi arcs of surface states}.

\begin{figure}[ht]

\includegraphics[width=0.99\columnwidth]{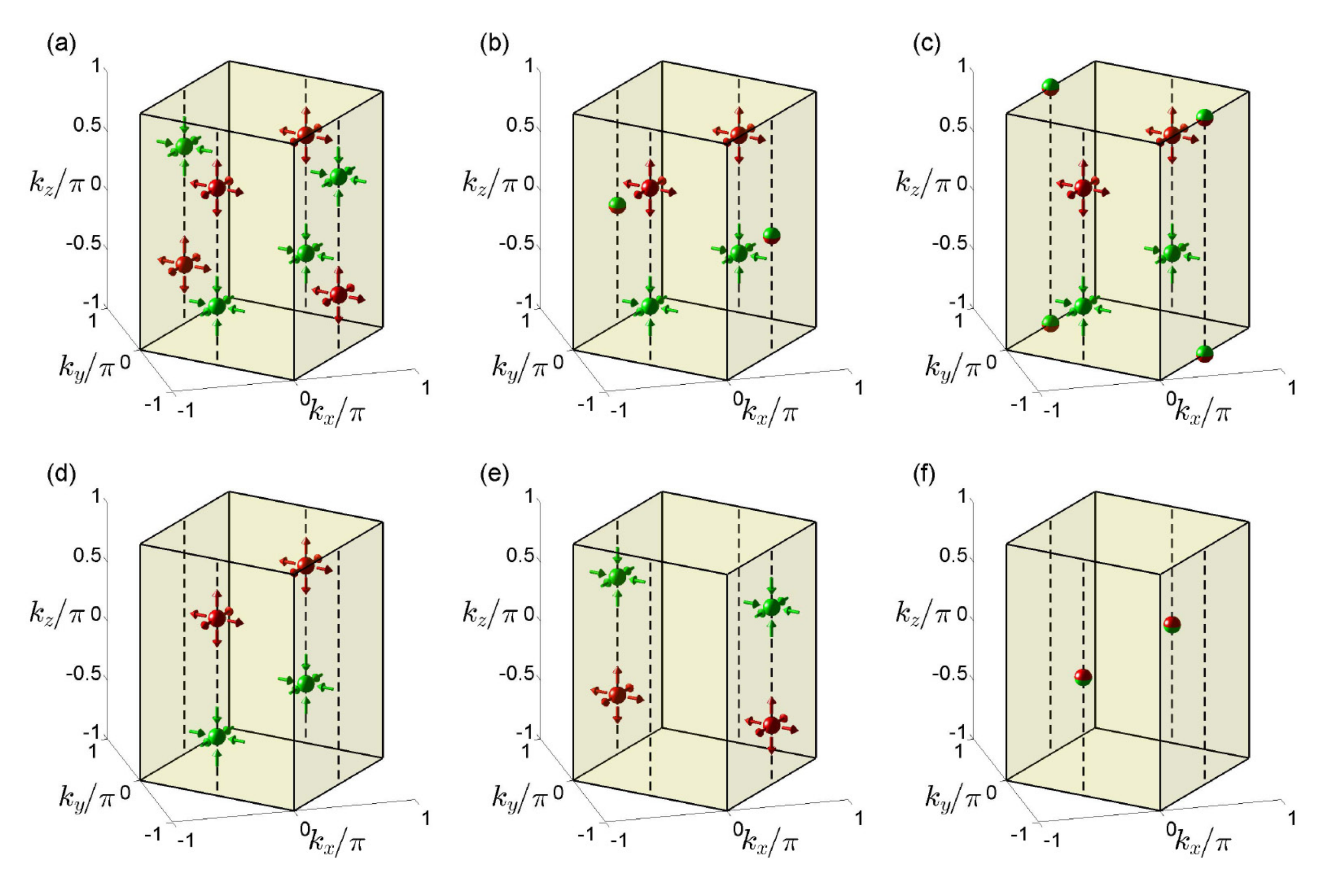}

\caption{   \textbf{ Weyl points in Weyl semimetal phases
for}
  (a) the WSM2 phase with $\alpha=0$ and $\beta=0$, (b) the boundary between the MSM2 phase  and the MSM1a phase with $\alpha=0.5$ and $\beta=-0.5$, (c) the boundary between  the MSM2 phase and the MSM1a phase with $\alpha=-0.5$ and $\beta=0.5$, (d) the MSM1a phase with $\alpha=0.8$ and $\beta=-0.8$,   (e) the MSM1b phase $\alpha=0.8$ and $\beta=0.8$,
   (f) the boundary between  the MSM1a phase and the band insulator phase with $\alpha=1.3$ and $\beta=-0.3$. For all of cases, we have set $t_x=t_y=t_z=t$. The yellow bulks represent the Brillouin zone; the red and green balls represent the Weyl points with positive and negative topological charges (also denoted by all-out and all-in arrows), respectively; the half red and half green balls represent merged Weyl points.
}\label{weyl}
\end{figure}
Here, we investigate moving and merging of Weyl points along with   varying of the dimensionless parameters $\alpha$ and $\beta$. Merging of Weyl points and annihilations of topological charges lead to topological phase transitions. In our model, there are four kinds of topological phase transitions such as (i) transition from the WSM2 phase to the MSM1a phase, (ii) transition from the MSM2 phase to the MSM1b phase,
 (iii) transition from the MSM1a phase to the band  insulator phase, and (iv) transition from the MSM1b phase to the band insulator phase.

 The WMS2 phase  has  two pairs of Weyl points $\mathbf{W}_{1,2}$ and $\mathbf{W}_{3,4}$ with topological charges $C_{1,2}=\pm 1$ and $C_{3,4}=\mp 1$, as shown in Fig.\ref{weyl}(a).
 When we keep $\alpha+\beta$ invariant and  increase $\alpha-\beta$,  the Weyl points $\mathbf{W}_{3,4}$ move towards each other and $\mathbf{W}_{1,2}$ stay at the original positions. When $\alpha-\beta$ increases to $1$, the Weyl points meet at $(\pi/2, -\pi/2, 0)$ in the Brillouin zone and merge, as shown  in Fig.\ref{weyl}(b). When $\alpha-\beta$ further  increases more than $1$, the Weyl points $\mathbf{W}_{3,4}$ annihilate and only
$\mathbf{W}_{1,2}$ remain, the system from the MSM2 phase   turns into the MSM1a phase, i.e.,   topological phase transition (i) happens. Topological phase transition (i) can also occur through the other type of moving and merging of Weyl points.  Starting from the WSM2 phase,    we keep $\alpha+\beta$ invariant and  decrease $\alpha-\beta$,  the Weyl points $\mathbf{W}_{3,4}$ move away from   each other and $\mathbf{W}_{1,2}$  stay at their starting positions. When $\alpha-\beta$ decreases to $-1$, the Weyl points $\mathbf{W}_{3,4}$ arrive at $(\pi/2, -\pi/2, \pm\pi)$, which are identical points in the Brillouin zone,  i.e., $\mathbf{W}_{3,4}$ meet and merge, as shown in Fig.\ref{weyl}(c). When $\alpha-\beta$ are less  than $-1$, $\mathbf{W}_{3,4}$ annihilate and topological phase transition (i)  happens. When it arrives at the MSM1a phase, there exist only  one pair of Weyl points $\mathbf{W}_{1,2}$, which have opposite topological charges, in the Brillouin zone, as shown  in Fig.\ref{weyl}(d).
Similarly, there are two types of moving and merging of Weyl points to realize topological phase transition (ii), i.e., the  transition from the WSM2 phase to the WSM1b phase.  We can vary the value of $\alpha+\beta$ and keep $\alpha-\beta$ invariant. When $\alpha+\beta$ increases to $1$ or $-1$, the Weyl points $\mathbf{W}_{1,2}$ meet and merge at the center or the  surface of the Brillouin zone, and $\mathbf{W}_{3,4}$ still remain. When $|\alpha+\beta|$ is greater than $1$, the Weyl points $\mathbf{W}_{1,2}$ annihilate and  a topological phase transition from the MSM2 phase into MSM1b phase happens, as shown in Fig.\ref{weyl}(e).
For the MSM1a phase, we can also   we increase $|\alpha+\beta|$ to $1$, the remaining Weyl points $\mathbf{W}_{1,2}$ meet and merge at the center or the surface of the Brillouin zone, as shown in Fig.\ref{weyl}(f). If we further increase $|\alpha+\beta|$ greater than $1$, the remaining Weyl points $\mathbf{W}_{1,2}$ annihilate and a gap opens, topological phase transition (iii) happens.
For the MSM1b phase, if we increase $|\alpha-\beta|$ to $1$, the remaining Weyl points $\mathbf{W}_{3,4}$ meet and merge at surface or corner of the Brillouin zone. If we further increase $|\alpha-\beta|$ greater than $1$, the remaining Weyl points $\mathbf{W}_{3,4}$ annihilate and a gap opens, so  topological phase transition (iv) happens.  In all the topological phase transitions, it is found that topological charges respect a conservation law and they are only created and annihilated in pairs.

\begin{figure}[ht]
\begin{center}
\includegraphics[width=0.99\columnwidth]{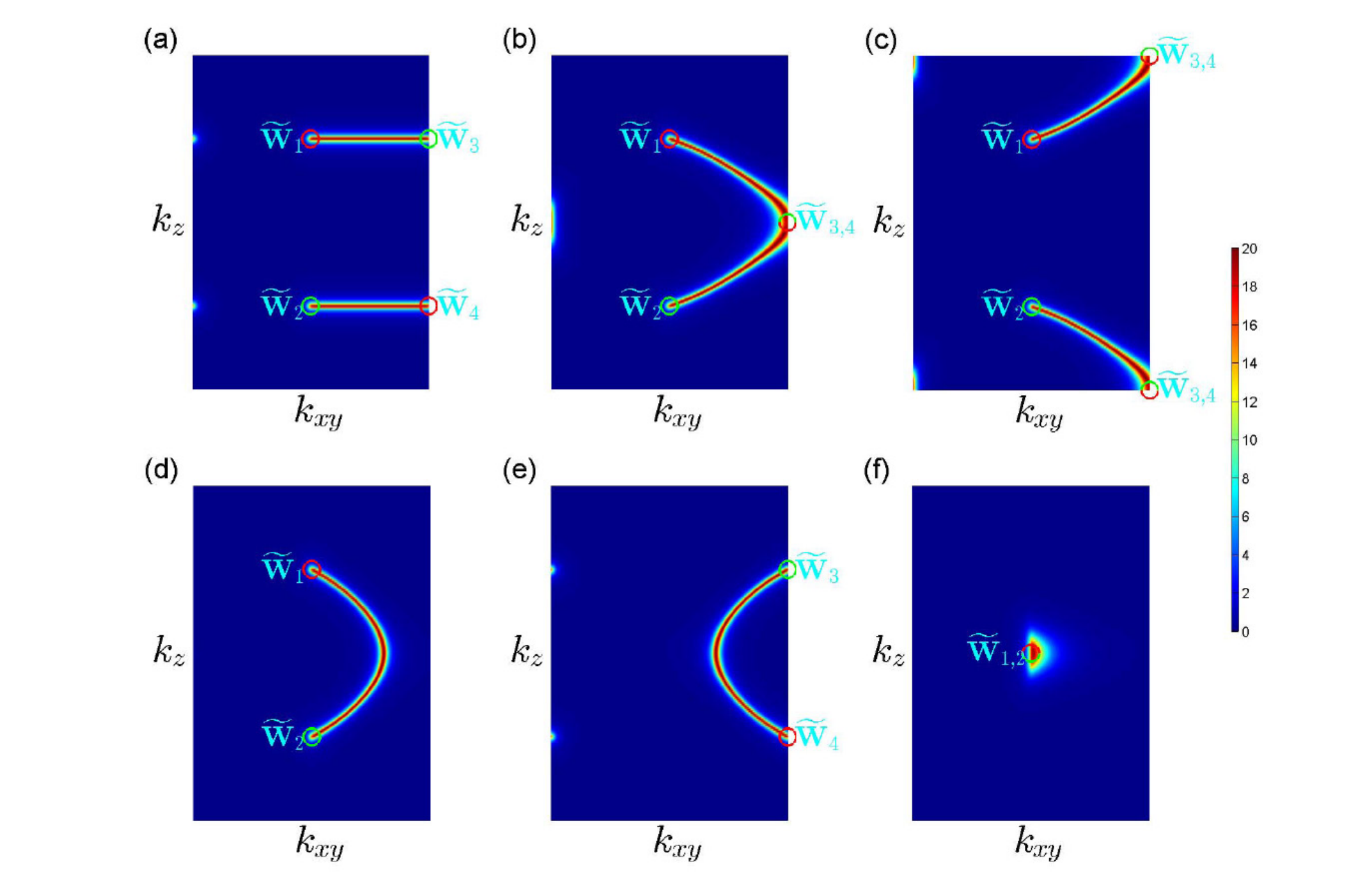}
\end{center}

\caption{   \textbf{ The spectral functions of surface  states with $E=0$ on the (010) surface Brillouin zone for}
  (a) the WSM2 phase with $\alpha=0$ and $\beta=0$, (b) the transition from the MSM2 phase to the MSM1a phase with $\alpha=0.5$ and $\beta=-0.5$, (c) the transition from the MSM2 phase to the MSM1a phase with $\alpha=-0.5$ and $\beta=0.5$, (d) the MSM1a phase with $\alpha=0.8$ and $\beta=-0.8$,   (e) the MSM1b phase $\alpha=0.8$ and $\beta=0.8$,
   (f) the transition from the MSM1a phase to the band insulator phase with $\alpha=1.3$ and $\beta=-0.3$. For all of cases, we have set $t_x=t_y=t_z=t$. Here, the rectangles represent the (010) surface Brillouin zone and $k_{xy}$ is the component on the $xy$ plane in momentum space for the wavevectors on the surface Brillouin zone. The red and green circles represent the projections of Weyl points with positive and negative topological charges on the surface Brillouin zone; the half red and half green circles represent the projections of merged Weyl points on  the surface Brillouin zone.
}\label{Fermiarc}
\end{figure}

In order to further study the characters of Weyl semimetals and topological phase transitions, we  calculate  the surface states of a slab geometry with (010) surfaces and investigate the evolution of  Fermi arcs along with the moving and merging of Weyl points. In Fig.\ref{Fermiarc}, we show the spectral function of the surface states at zero energy. The spectral function can be calculated through the formula $A(E)=-\frac{1}{\pi}\textrm{Im}G^r(E)$, where $G^r(E)$ is the retarded Green function of the system. The projections of Weyl points $\mathbf{W}_{1,2,3,4}$  on the surface Brillouin zone are denoted as
$\widetilde{\mathbf{W}}_{1,2,3,4}$. Fig.\ref{Fermiarc}(a) shows that, in the WSM2 phase, there are two Fermi arcs in the surface Brilllouin zone, which connect points $\widetilde{\mathbf{W}}_{1}$ and $\widetilde{\mathbf{W}}_{3}$,  $\widetilde{\mathbf{W}}_{2}$ and $\widetilde{\mathbf{W}}_{4}$, respectively. Since Weyl points $\mathbf{W}_1$ and $\mathbf{W}_3$, $\mathbf{W}_2$ and $\mathbf{W}_4$ have opposite topological charges, we conclude that Fermi arcs connect the projections of Weyl points with opposite topological charges on  the surface Brillouin zone. When $|\alpha-\beta|$ increases to $1$,  $\widetilde{\mathbf{W}}_3$ and $\widetilde{\mathbf{W}}_4$ meet and merge at the side boundary or the corners of the surface Brillouin zone, thereby two Fermi arcs combine into one single Fermi arc, as shown in Fig.\ref{Fermiarc}(b) and (c), which corresponds to topological phase transition (i). When $|\alpha-\beta|$ increases greater than $1$, the system in the MSM1a phase, the Fermi arc connects the projections of the remaining Weyl points  $\widetilde{\mathbf{W}}_{1,2}$,  as shown in Fig.\ref{Fermiarc}(d). Similarly, for the MSM1b phase, there exists a Fermi arc connect the points
$\widetilde{\mathbf{W}}_3$ and $\widetilde{\mathbf{W}}_4$ in the surface Brillouin zone,  as shown in Fig.\ref{Fermiarc}(e). When the transition from the MSM1a phase or the MSM1b phase to the band insulator phase the happen, the Fermi arc firstly shrink into a point, as shown in Fig.\ref{Fermiarc}(f), and finally disappears.

\vspace{0.5cm} \noindent \textbf{Hidden symmetry at Weyl points}.

 Here, we build the hidden symmetry at Weyl points. For convenience to construct the hidden symmetry, we suppose the case with the Hamiltonian $H_0$ as Eq.(\ref{tbh}) as the original model and the total model $H=H_0+H_d+H_s$ as the modified model.

\vspace{0.3cm} \noindent\textit{Hidden symmetry at Weyl points of
the original model}. In the following, we will show that the Weyl
points in the original model are protected by   a HAS. For the
original model, the lattice is invariant under the operation defined
as
\begin{eqnarray}
\Upsilon=(e^{i\pi})^{i_z}\sigma_xT_{\hat{x}}K,
\end{eqnarray}
 where $K$  is the
complex conjugate operator; $T_{\hat{x}}$ is a translation operator
that moves the lattice along the $x$ direction by a unit vector;
$\sigma_x$ is the Pauli matrix representing the sublattice exchange;
$(e^{i\pi})^{i_z}$ is a local $U(1)$ gauge transformation.  It is
easy to prove that the symmetry operator $\Upsilon$ is antiunitary,
and its square is equal to $\Upsilon^2=T_{2\hat{x}}$.

By setting $\alpha=0$ and $\beta=0$, the Bloch Hamiltonian of original model can be obtain from Eq.(\ref{BH1}) as
\begin{eqnarray}
\mathcal{H}_0(\mathbf{k})&=&-2t_x\cos k_x \sigma_x-2t_y\cos k_y \sigma_y -2t_z \cos k_z \sigma_z. \label{BH0}
\end{eqnarray}
  The  symmetry operator $\Upsilon$ can be
considered as a self-mapping of  the original model defined as
\begin{eqnarray*}
\Upsilon:(\mathbf{k}, {\cal H}_0(\mathbf{k}),
\Psi_{0,\mathbf{k}}(\mathbf{r}))\mapsto (\mathbf{k}', {\cal
H}_0(\mathbf{k}'), \Psi_{0,\mathbf{k}'}'(\mathbf{r})),
\end{eqnarray*}
where $\Psi_{0,\mathbf{k}}(\mathbf{r})$ and
$\Psi'_{0,\mathbf{k}'}(\mathbf{r})$ are the Bloch functions of the
original model. We suppose that the Bloch functions of the square
lattice model have the form as
\begin{eqnarray}
\Psi_{0,\mathbf{k}}=\left(\matrix{u^{s}_{1,\mathbf{k}}(\mathbf{r})\cr
u^{s}_{2,\mathbf{k}}(\mathbf{r})}\right)e^{i\mathbf{k}\cdot\mathbf{r}},
\label{Bloch}
\end{eqnarray}
with
$u^s_{i,\mathbf{k}}(\mathbf{r})=u^s_{i,\mathbf{k}}(\mathbf{r}+\mathbf{R}_n)$
with $i=1,2$ for two sublattices and $\mathbf{R}_n$ being a lattice
vector. Performing the   symmetry transformation on the Bloch
function (\ref{Bloch})
 leads to
$\Upsilon\Psi_{0,\mathbf{k}}(\mathbf{r})={\Psi'_{0,\mathbf{k}'}}(\mathbf{r})$
with $ \mathbf{k}'= (k_x', k_y', k_z')=(-k_x, -k_y,  -k_z+\pi)$. If
the condition $\mathbf{k}'=\mathbf{k}+\mathbf{G}$, where
$\mathbf{G}$ is the reciprocal lattice vector, is satisfied,
$\mathbf{k}$ is a $\Upsilon$-invariant point.  In the Brillouin
zone, the distinct $\Upsilon$-invariant points are
$\mathbf{M}_{1,2}=(\pi/2,\pi/2, \pm\pi/2)$,
$\mathbf{M}_{3,4}=(\pi/2,-\pi/2, \pm\pi/2)$,
$\mathbf{N}_{1,2}=(0,0,\pm\pi/2)$, and $\mathbf{N}_{3,4}= (0,   \pi,
\pm\pi/2) $. The square of the $\Upsilon$ operator can be written in
the form as $\Upsilon^2=T_{2\hat{x}}=e^{-2ik_x}$ in the Bloch
representation. It is easy to verify that  $\Upsilon^2=-1$ the
points $\mathbf{M}_i(i=1,2,3,4)$, while $\Upsilon^2=1$ at the points
$\mathbf{N}_i (i=1,2,3,4)$.  Considering the antiunitarity of the
operator $\Upsilon$, based on Kramers  theorem,  we can
conclude that there must be band degeneracies at the
$\Upsilon$-invariant points $\mathbf{M}_i (i=1,2,3,4)$, which are
just the Weyl points $\mathbf{W}_i (i=1,2,3,4)$ in the MSM2 phase with $\alpha=0$ and $\beta=0$. There, a hidden symmetry with its square of operators being $-1$ exists at the Weyl points of the
original model.

\vspace{0.3cm} \noindent\textit{Hidden symmetry at Weyl points of
the modified model}.
  It is easy to verify that the HAS
$\Upsilon$ is violated in the modified model. However, with the
mapping $\Omega_{\alpha,\beta}$ from the modified model into the
original model defined in section Methods, we can find the HAS in
the modified model. Based on the mapping $\Omega_{\alpha,\beta}$, we
define an operation
$\Lambda_{\alpha,\beta}=\Omega_{\alpha,\beta}^{-1}\circ\Upsilon\circ\Omega_{\alpha,\beta}$,
which can be regarded as a self-mapping of the modified model as
\begin{eqnarray*}
\Lambda_{\alpha,\beta}:(\mathbf{k},\mathcal{H}(\mathbf{k}),\Psi_{\mathbf{k}}(\mathbf{r}))
\mapsto
(\mathbf{k}',\mathcal{H}(\mathbf{k}'),\Psi'_{\mathbf{k}'}(\mathbf{r}))
\end{eqnarray*}
Performing the above operation on the Bloch function of the modified
model, we have
$\Lambda_{\alpha,\beta}\Psi_{\mathbf{k}}(\mathbf{r})=\Psi_{\mathbf{k}'}(\mathbf{r})$,
where
$\mathbf{k}'=(-k_x,-k_y,-k_z-\Delta_z(\mathbf{k})-\Delta_z(\mathbf{k}')+\pi)$
with $\Delta_z(\mathbf{k})=K_z-k_z$ being the shift   of the
$z$-component of the wave vector $\mathbf{k}$ due to the mapping
  $\Omega_{\alpha,\beta}$. If $\mathbf{k}'=\mathbf{k}+\mathbf{G}$ is
  satisfied, $\mathbf{k}$ is a $\Lambda_{\alpha,\beta}$-invariant point.
In the Brillouin zone, the distinct
$\Lambda_{\alpha,\beta}$-invariant points are
$\mathbf{P}_{1,2}=(\pi/2,\pi/2,\pm \arccos(\alpha+\beta))$,
$\mathbf{P}_{3,4}=(\pi/2,-\pi/2,\pm\arccos(\alpha-\beta))$,
$\mathbf{Q}_{1,2}=(0,0,\pm\arccos \alpha)$ and
$\mathbf{Q}_{3,4}=(0,\pi,\pm\arccos \alpha)$.

From the definition of the operator $\Lambda_{\alpha,\beta}$, we can
verify
$\Lambda_{\alpha,\beta}^2=\Omega_{\alpha,\beta}^{-1}\circ\Upsilon^2\circ\Omega_{\alpha,\beta}$,
which acts on the Bloch function as
$\Lambda_{\theta,\beta}^2\Psi_{\mathbf{k}}(\mathbf{r})=e^{-2i k_x
}\Psi_{\mathbf{k}}(\mathbf{r}) $. Substituting  the
$\Lambda_{\alpha,\beta}$-invariant points $\mathbf{P}_{i}
(i=1,2,3,4)$ and $\mathbf{Q}_i (i=1,2,3,4)$ into the above equation,
we find $\Lambda_{\alpha,\beta}^2=-1$ at $\mathbf{P}_i (i=1,2,3,4)$,
while $\Lambda_{\alpha,\beta}^2= 1$ at $\mathbf{Q}_i (i=1,2,3,4)$.
Since $\Lambda_{\alpha,\beta}$ is an antiunitary operator, based on Kramers
  theorem,  there must exist band degeneracies  at the
$\Lambda_{\alpha,\beta} $-invariant points $\mathbf{P}_{i}
(i=1,2,3,4)$, which  are just the Weyl points $\mathbf{W}_{i}
(i=1,2,3,4)$ of the WSM2 phase for $|\alpha+\beta|<1$ and
$|\alpha-\beta|<1$. For $|\alpha+\beta|<1$ and $|\alpha-\beta|>1$,
$\mathbf{P}_{3,4}$ do not exist, there are only  the
$\Lambda_{\alpha,\beta} $-invariant points $\mathbf{P}_{1,2}$, which
correspond  to the Weyl points $\mathbf{W}_{1,2}$ of the WSM1a
phase. Similarly, for $|\alpha+\beta|>1$ and $|\alpha-\beta|<1$,
$\mathbf{P}_{1,2}$ do not exist, there are only  the
$\Lambda_{\alpha,\beta} $-invariant points $\mathbf{P}_{3,4}$, which
correspond  to the Weyl points $\mathbf{W}_{3,4}$ of the WSM1b
phase. For the case $|\alpha+\beta|>1$ and $|\alpha-\beta|>1$, all
the $\Lambda_{\alpha,\beta} $-invariant points
$\mathbf{P}_{1,2,3,4}$ do not exist, so there are not  Weyl points,
which corresponds to a band insulator phase.

 \begin{figure}[th]

\includegraphics[width=0.99\columnwidth]{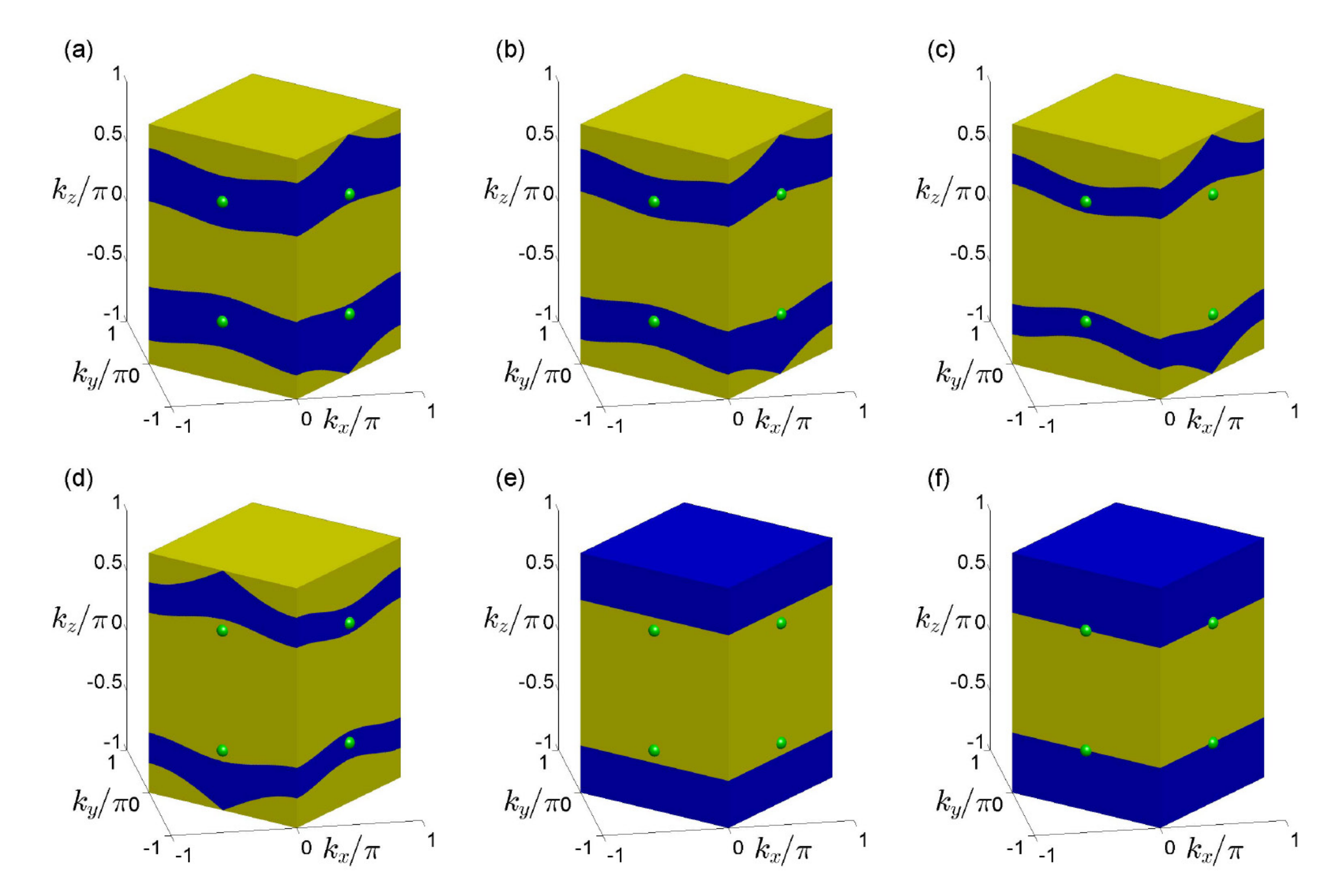}

\caption{  \textbf{The mapping from the Brillouin zone of the
modified model into the Brillouin zone of the original model for}
(a) $\alpha=-0.3, \beta=0.3$; (b) $\alpha=1, \beta=-0.8$;
(c)$\alpha=1, \beta=0.8$; (d) $\alpha=2, \beta=0$; (e) $\alpha= 0.6,
\beta=-0.4$; (f) $\alpha=1, \beta=0$; . Here, the yellow bulk
represents the Brillouin zone of the original model; the cyan balls
mark the Weyl points of the original model; the blue part represent
the image of the mapping in the Brillouin zone of the original model
for the Brillouin zone of the modified model. }\label{fig5}
\end{figure}

We can interpret the above results in an intuitive way. The mapping
$\Omega_{\alpha,\beta}$ from the Brillouin zone of the modified
model into that of the original model is not surjective, which can
be seen in Fig.\ref{fig5}.  For the WSM2 phase, i.e.
$|\alpha+\beta|<1$ and $|\alpha-\beta|<1$, the image of the mapping
for the Brillouin zone of the modified model covers the degenerate
$\Upsilon$-invariant points $\mathbf{M}_{1,2}$ and
$\mathbf{M}_{3,4}$ in the Brillouin zone of the original model, as
shown in Fig.\ref{fig5}(a). Therefore, there are always two pairs of
$\Lambda_{\alpha,\beta} $-invariant points $\mathbf{P}_{1,2}$ and
$\mathbf{P}_{3,4}$, where  the Weyl points locate, map into the
degenerate $\Upsilon$-invariant points $\mathbf{M}_{1,2}$ and
$\mathbf{M}_{3,4}$. When we increase $|\alpha-\beta|$ to $1$,
$\mathbf{P}_3$ and $\mathbf{P}_4$ become the same point of the
Brillouin zone of the modified model, which maps into the points
$\mathbf{M}_{3,4}$
 in the Brillouin zone of the original model as shown in Fig.\ref{fig5}(b), so  the corresponding two  Weyl points
 merge and a phase transition  from the WSM2 phase to the WSM1a phase occurs.
   When
 $|\alpha+\beta|<1$ and $|\alpha-\beta|>1$,
  the image of mapping for
 the Brillouin zone of the modified model only covers the degenerate
 $\Upsilon$-invariant points $\mathbf{M}_{1,2}$ in the Brillouin zone of the  original
 model as shown in
 Fig.\ref{fig5}(c). There exists
 a pair of Weyl points $\mathbf{P}_{1,2}$ in the Brillouin zone of the modified model
 mapping into the degenerate $\Upsilon$-invariant points
 $\mathbf{M}_{1,2}$, which
 corresponds to the WSM1a phase.  Similarly, when
 $|\alpha+\beta|=1$ and $|\alpha-\beta|<1$, the image of the mapping
 for the Brillouin zone of the modified model covers $\mathbf{M}_{1,2}$ and
 $\mathbf{M}_{3,4}$ while $\mathbf{M}_{1,2}$ locate at the edge of
 the image and the same point in the Brillouin zone of the modified model maps into $\mathbf{M}_{1,2}$. The   degenerate $\Lambda_{\alpha,\beta}
 $-invariant points
 $\mathbf{P}_{1,2}$ merge  at the edge of the Brillouin zone   of the modified model while
   $\mathbf{P}_{3.4}$ still exist. This case   corresponds to the phase boundary between the WSM2
 and WSM1b phases.  When $|\alpha+\beta|>1$ and
 $|\alpha-\beta|<1$, the image of the mapping for the Brillouin zone of the
 modified model only covers the degenerate $\Upsilon$-invariant
 points $\mathbf{M}_{3,4}$ as shown in Fig.\ref{fig5}(d).
 Correspondingly, there exists a pair of Weyl points
 $\mathbf{P}_{3,4}$ in the Brillouin zone of the modified model mapping into the
 degenerate $\Upsilon$-invariant points $\mathbf{M}_{3,4}$, which
 corresponds to the WSM1b phase.
 When $|\alpha+\beta|>1$ and $|\alpha-\beta|>1$,  the image of mapping for the Brillouin zone of the
 modified model does not cover any degenerate $\Upsilon$-invariant
 points in the Brillouin zone of the original model as shown in Fig.\ref{fig5}(e).
 Therefore, there does not exist any Weyl point in the Brillouin zone of
 the modified model and a gap opens, which corresponds to the band
 insulator phase.
When $|\alpha+\beta|=1$ and $|\alpha-\beta|=1$, a direct phase
transition between the MSM2 phase and the band
 insulator phase can occurs, where two pairs of Weyl points merge
 simultaneously. For this case,
  the edge of the image of the mapping for the Brillouin zone of the modified model
 covers the degenerate $\Upsilon$-invariant points
 $\mathbf{M}_{1,2}$ and $\mathbf{M}_{3,4}$ in the Brillouin zone of the
 original model as shown in Fig.\ref{fig5}(f), which means that the four $\Lambda_{\alpha,\beta} $-invariant $\mathbf{P}_{1,2}$ and $\mathbf{P}_{3,4}$ merge as two points. Therefore, the two pairs
 of Weyl points simultaneously merge at  the edge of the Brillouin zone of
 the modified model.

\vspace{0.5cm} \noindent \textbf{Discussion}.

  In summary, we have proposed a scheme to realize Weyl semimetals in a cubic optical lattice. There exist three Weyl semimetal phases, such as the WSM2, WSM1a, and WSM1b phases, for different parameter ranges. In the Brillouin zone, there are two pairs of Weyl points for the WSM2 phase while there is one pair of Weyl points for the MSM1a and MSM1b phases. The Weyl points move along with varying of the parameters. When the Weyl  points with opposite topological charges meet, they merge and annihilate, which leads to a  topological phase transition. The spectral functions of surface states at zero energy for a slab with  (010) surfaces have been calculated. Fermi arcs appear to connect the projection of the Weyl points with opposite topological charges on the surface Brillouin zone.  There are two Fermi arcs in the WSM2 phase and there is one in the MSM1a and MSM1b phases. When the phase transition from the WSM2 phase to the MSM1a or MSM1b phase happens, the two Fermi arcs combine into one Fermi arc.
For the phase transition from the MSM1a or MSM1b  phase to the band insulator phase, the Fermi arc shrink into a point, then disappears. We also found that there exist hidden symmetries at all of Weyl points. These hidden symmetries have an antiunitary operator with its square being $-1$. Based on the mapping method\cite{Hou2}, we constructed hidden symmetries at all of Weyl points. Our work deepens our understanding of Weyl semimetals on the point view of symmetry.

\vspace{0.5cm} \noindent \textbf{Methods}.

\noindent \textbf{The mapping from the modified model into the
original model}.
 We can define a mapping from the modified
model into the original model as\cite{Hou2}
\begin{eqnarray*}
\Omega_{\alpha,\beta}:&& (\mathbf{k}, \mathcal{H} (\mathbf{k}),
\Psi_{\mathbf{k}}(\mathbf{r}))\mapsto  (\mathbf{K},
\mathcal{H}_0(\mathbf{K}), \Psi_{0,\mathbf{K}}(\mathbf{r}))
\end{eqnarray*}
where $\Psi_{ \mathbf{k}}(\mathbf{r})$ represents the Bloch
function of the modified model. The concrete  form of the mapping
$\Omega_{\alpha,\beta}$ depends on the dimensionless parameters with
$\alpha$ and $\beta$.
 For this mapping, we have
$\Omega_{\alpha,\beta}\Psi_{ \mathbf{k}}(\mathbf{r})=\Psi_{0,\mathbf{K}}(\mathbf{r})$
with
\begin{eqnarray}
K_x&=&k_x\label{KTX}\\
K_y&=&k_y\label{KTY}\\
 K_z &=&\left\{\matrix{-\mathcal{K}_{\alpha,\beta}(\mathbf{k}), & k_z\in[-\pi, 0]\cr
 \mathcal{K}_{\alpha,\beta}(\mathbf{k}), & k_z\in[0,
 \pi]}\right.,\label{KTZ}
 \end{eqnarray}
 where $\mathcal{K}_{\alpha,\beta}(\mathbf{k})\equiv \arccos\left(\frac{ \cos k_z-\alpha- \beta\sin k_x \sin
 k_y}{1+|\alpha|+|\beta|}\right)$.
 Replacing $\mathbf{k}$ in equation (\ref{BH1}) with $\mathbf{K}$ via equations
 (\ref{KTX}), (\ref{KTY}), and (\ref{KTZ}), we obtain
\begin{eqnarray}
\mathcal{H}_0(\mathbf{K})&=&-2t_x'\cos K_x \sigma_x- 2t_y'\cos
K_y\sigma_y -2t_z'\cos K_z\sigma_z
\end{eqnarray}
with $t_x'=t_x$, $t_y'=t_y$, and $t_z'=t_z+2|t_{xy}|+|v|/2$, which
is just the Bloch Hamiltonian (\ref{BH0}) of the original model.

\section*{Acknowledgement}
\noindent
   J.M.H. acknowledges the support from the National Natural Science Foundation of China under Grant
No. 11274061; W.C. acknowledges the supports from the National Natural Science Foundation of China
under Grant No. 11504171, the Natural Science Foundation of Jiangsu Province, China under Grants
No. BK20150734, and the Project funded by China Postdoctoral Science Foundation under Grants No.
2014M560419 and No. 2015T80544.

\section*{Author contributions}
\noindent J.M.H conceived and supervised the project. J.M.H. and
W.C. made the calculations and wrote the paper.

\section*{Additional Information}

\noindent{\bf Competing financial interests:}  The authors declare
no competing financial interests.


\section*{References}
\begin{thebibliography}{99}
\bibitem{Hasan10} Hasan, M. Z. \& Kane, C. L. Topological insulators. \em Rev. Mod. Phys. \em \textbf{82}, 3045 (2010).

\bibitem{Qi11} Qi, X. L. \& Zhang, S. C. Topological insulators and superconductors. \em Rev. Mod. Phys. \em \textbf{83}, 1057 (2011).

\bibitem{Thouless} Thouless, D. J., Kohmoto, M., Nightingale, M. P. \& Nijs, M. D. Quantized Hall conductance in a two-dimensional perioidc potential. \em Phys. Rev. Lett. \em \textbf{49}, 405 (1982).

\bibitem{Haldane} Haldane F. D. M. Model for a quantum Hall effect without Landau levels: condensed-matter realization of the ``partiy anomaly''. \em Phys. Rev. Lett. \em \textbf{61}, 2015 (1988).

\bibitem{Kane} Kane, C. L. \& Mele, E. J. Quantum  spin Hall effect in graphene. \em Phys. Rev. Lett. \em \textbf{95}, 226801 (2005).

\bibitem{Read} Read, N. \& Green, D. Paired states  of fermions in two dimensions with breaking of parity and time-reversal symmetries and the fractional quantum Hall effect. \em Phys. Rev. B \em \textbf{61}, 10267 (2000).

\bibitem{Qi2} Qi, X. L., Hughes, T. L., Raghu, S. \& Zhang, S. C. Time-reversal-invariant topological superconductors and superfluids in two and three dimensions \em Phys. Rev. Lett. \em \textbf{102}, 187001 (2009).

\bibitem{Wan}Wan, X., Turner, A. M., Vishwanath, A.  \& Savrasov, S. Y.  Topological semimetal and
Fermi-arc surface states in the electronic structure of pyrochlore
iridates. \em Phys. Rev. B \em \textbf{83},  205101 (2011).

\bibitem{Xu}Xu, G., Weng, H., Wang, Z., Dai, X.  \& Fang, Z. Chern semimetal and the quantized anomalous Hall effect in HgCr$_2$Se$_4$. \em Phys.
Rev. Lett. \em \textbf{107},  186806 (2011).

\bibitem{Burkov1}Burkov, A. A., Hook, M. D. \& Balents, L.  Topological nodal semimetals. \em Phys. Rev.
B \em \textbf{84},  235126 (2011).

\bibitem{Burkov2}  Burkov, A. A. \& Balents, L.  Weyl semimetal in a topological insulator multilayer. \em Phys. Rev.
Lett. \em \textbf{107},  127205 (2011).

\bibitem{Zyuzin}Zyuzin, A. A., Wu, S. \& Burkov, A. A.  Weyl semimetal with broken time reversal and inversion symmetries. \em Phys.
Rev. B \em \textbf{85}  165110 (2012).


\bibitem{Yang14} Yang, B. J. \& Nagaosa, N. Classification of stable three-dimensional Dirac semimetals with nontrivial topology. \em Nat. Commun. \em \textbf{5} 4898 (2014).

\bibitem{Xu1}Xu, S. Y.  \emph{et al}  Observation of Fermi arc surface states in a topological metal. \em Science \em
\textbf{347}, 294 (2015). 


\bibitem{Weng}Weng, H., Fang, C., Fang, Z., Bernevig, B.  A. \& Dai,
X.  Weyl semimetal phase in noncentrosymmetric transition-metal
monophosphides. \em Phys. Rev. X \em \textbf{5},  011029
  (2015).

\bibitem{Huang} Huang, S. M.  \emph{et al}.  An inversion breaking Weyl semimetal state in the TaAs material class. \em Nat. Commun. \em
\textbf{6}, 7373 (2015).

\bibitem{Xu2} Xu, S. Y.  \emph{et al}.  Discovery of a Weyl fermion semimetal and topological Fermi arcs. \em Science \em \textbf{349}, 613 (2015).

\bibitem{Lv} Lv, B.  Q.  \emph{et al}.  Experimental discovery of Weyl semimetal TaAs. \em Phys. Rev. X \em \textbf{5} 031013 (2015).

\bibitem{Xu3} Xu, S. Y.  \emph{et al}. Discovery of a Weyl fermion state with Fermi arcs in niobium arsenide. \em Nat. Phys. \em \textbf{11}, 748 (2015).





\bibitem{Ganeshan15} Ganeshan, S. \& Das Sarma, S. Constructing a Weyl semimetal by stacking one-dimensional topological phases. \em Phys. Rev. B \em \textbf{91}, 125438 (2015).


\bibitem{Zhu} Zhu, S. L., Wang, B. \& Duan, L. M. Simulation and
dectection of Dirac fermions with cold atoms in an optical lattice.
\em Phys. Rev. Lett. \em \textbf{98}, 260402 (2007).

\bibitem{Wunsch} Wunsch, B., Guinea, F. \& Sols, F. Dirac-point
engineering and topological phase transitions in honeycomb optical
lattices. \em New J. Phys. \em \textbf{10}, 103027 (2008).



 %

\bibitem{Hou09} Hou, J. M., Yang, W. X. \& Liu X. J. Massless Dirac fermions in a square optical lattice. \em Phys. Rev. A \em \textbf{79}, 043621 (2009).

\bibitem{Hou1}Hou, J. M.  Hidden-symmetry-protected topological semimetals on a square lattice. \em Phys. Rev. Lett. \em \textbf{111},  130403
(2013).


\bibitem{Hou2}Hou, J. M.  Moving and merging of Dirac points on a square lattice and hidden symmetry protection. \em Phys. Rev. B \em
\textbf{89},
235405 (2014).

\bibitem{Tarruell} Tarruell, L., Greif, D., Uehlinger, T. \&
Esslinger, T. Creating, moving and merging Dirac points with a Fermi
gas in a tunable honeycomb lattice. \em Nature \em \textbf{483},
302-305 (2012).



\bibitem{Delplace}Delplace, P., Li, J.  \& Carpentier, D.  Topological Weyl semi-metal from a lattice model. \em Europhys.
Lett. \em \textbf{97},  67004 (2012).

\bibitem{Jiang}Jiang, J. H. Tunable topological Weyl semimetal from simple-cubic lattices with staggered fluxes. \em Phys. Rev. A \em
\textbf{85}, 033640 (2012).

\bibitem{Dubcek} Dub\v{c}ek, T. \emph{et al}. Weyl points in three-dimensional optical lattices: synthetic magnetic monopoles in momentum space. \em Phys. Rev. Lett. \em \textbf{114}, 225301 (2015).

\bibitem{XuY} Xu, Y. \& Zhang, C. Dirac and Weyl rings in three dimensional cold atom optical lattices. \em Phys. Rev. A \em \textbf{93}, 063606 (2016).

\bibitem{Struck12} Struck, J. \emph{et al}. Tunable gauge potential
for neutral and spinless particles in driven optical lattices. \em Phys. Rev. Lett. \em \textbf{108}, 225304 (2012).

\bibitem{Struck13} Struck, J. \emph{et al}.
Engineering Ising-XY spin models in a triangular lattice using tunnable artificial gauge fields. \em Nat. Phys. \em \textbf{9}, 738 (2013).

\bibitem{Aidelsburger11} Aidelsburger, M. \emph{et al}. Experimental realization of strong effective magnetic fields in an opticl lattice. \em Phys. Rev. Lett. \em \textbf{107}, 255301 (2011).

\bibitem{Miyake13} Miyake, H. \emph{et al}. Realizing the Harper Hamiltonian with laser-assisted tunneling in optical lattices.
\em Phys. Rev. Lett. \em \textbf{111}, 185302 (2013).

\bibitem{Aidelsburger13} Aidelsburger, M. \emph{et al}. Realization of  the Hofstadter Hamiltonian with ultracold atoms in optical lattices. \em Phys. Rev. Lett. \em \textbf{111}, 185301 (2013).



\bibitem{Jimenez-Garcia12} Jim\'enez-Garc\'ia, K. \emph{et al}. Peierls substitution in an engineered lattice
potential. \em Phys. Rev. Lett. \em \textbf{108}, 225303 (2012).

\bibitem{Atala13} Atala, M. \emph{et al}. Direct measurement of the Zak phase in topological Bloch bands. \em  Nat. Phys.  \em \textbf{9}, 795 (2013).

 




\end{thebibliography}
\end{document}